\documentclass[nofootinbib,pre,showpacs,showkeys,onecolumn,groupaddress,preprintnumbers,floatfix]{revtex4-1}

\usepackage{dynlearn}

\newcommand{\argmin}{\text{argmin}}

\newcommand{\Shannon}{H}
\newcommand{\kB}{k_\text{B}}
\newcommand{\Wdiss}{W_\text{diss}}
\newcommand{\MI}{\text{\textbf{I}}}
\newcommand{\TC}{C_\text{tot}}

\newcommand{\St}{\mathcal{S}}
\newcommand{\MSt}{\mathcal{M}}
\newcommand{\SSet}{\boldsymbol{\mathcal{S}}}

\newcommand{\MSet}{\boldsymbol{\mathcal{M}}}

\newcommand{\drivinghistory}[1][t]{\overleftarrow{x}_{#1}}
\newcommand{\drive}{x_{0:\tau}}
\renewcommand{\T}{\mathsf{T}}
\newcommand{\Transition}{\T_{\drive}}
\newcommand{\GlobalEq}[1][x_0]{\boldsymbol{\pi}_{#1}}
\newcommand{\LocalEq}[2][x_0]{\GlobalEq[#1]^{(#2)}}
\newcommand{\LocalZ}[2][x_0]{Z_{#1}^{(#2)}}
\newcommand{\LocalFEq}[2][x_0]{F_{#1}^{(#2)}}
\newcommand{\LocalFadd}[2][t]{F_{{\drivinghistory[#1]}, \text{add}}^{(#2)}}

\newcommand{\actual}[1][0]{\boldsymbol{\mu}_{#1}}
\newcommand{\asif}[1][0]{\boldsymbol{q}_{#1}}

\newcommand{\ModularityDiss}{\braket{\Wdiss^{(\text{mod})}}}
\newcommand{\MismatchDiss}{\braket{\Wdiss^{(\text{mismatch})}}}

\newcommand{\zero}{{\color{blue}{\boldsymbol{\mathtt{0}}}}}
\renewcommand{\one}{{\color{blue}{\boldsymbol{\mathtt{1}}}}}

\newcommand{\autocite}{\cite}
\renewcommand{\DKL}{D_\text{KL}}

\newenvironment{absolutelynopagebreak}
  {\par\nobreak\vfil\penalty0\vfilneg
   \vtop\bgroup}
  {\par\xdef\tpd{\the\prevdepth}\egroup
   \prevdepth=\tpd}

\def\tbf #1 {\textbf{#1} }

\begin{document}

\def\ourTitle{Transforming Metastable Memories:\\The 
    Nonequilibrium Thermodynamics of Computation
}

\def\ourAbstract{Framing computation as the transformation of metastable memories, we explore its fundamental thermodynamic limits.  
The true power of information follows from a novel decomposition of nonequilibrium free energy derived here, 
which provides a rigorous thermodynamic description of coarse-grained memory systems.  
In the nearly-quasistatic limit, logically irreversible operations can be performed with thermodynamic reversibility.  
Yet, here we show that beyond the reversible work Landauer's bound requires of computation, 
dissipation must be incurred both for modular computation and for neglected statistical structure among memory elements used in a computation.  
The general results are then applied to evaluate the thermodynamic costs of all two-input--one-output logic gates, 
including the universal NAND gate.  Interwoven discussion clarifies the prospects for Maxwellian demons and information engines 
as well as opportunities for hyper-efficient computers of the future.
}

\def\ourKeywords{  nonequilibrium thermodynamics, computation, Landauer bound, mismatch dissipation, modularity
}

\hypersetup{
  pdfauthor={Paul M. Riechers},
  pdftitle={\ourTitle},
  pdfsubject={\ourAbstract},
  pdfkeywords={\ourKeywords},
  pdfproducer={},
  pdfcreator={}
}

\title{\ourTitle}

\author{Paul M. Riechers}
\email{pmriechers@ucdavis.edu}

\affiliation{Complexity Sciences Center and Physics Department,
University of California at Davis, One Shields Avenue, Davis, CA 95616}

\date{\today}
\bibliographystyle{unsrt}

\begin{abstract}
\ourAbstract
\end{abstract}

\keywords{\ourKeywords}

\pacs{
89.70.+c  02.50.Ey  02.50.-r  05.20.-y  }

\date{\today}

\begin{absolutelynopagebreak}
\maketitle

\tableofcontents

\end{absolutelynopagebreak}

\setstretch{1.1}

\pagebreak

\section{Introduction}

	Modern scientific understanding suggests that computation can be performed
	without any dissipation 
	at all---a perplexing result since we still
	plug in our computers 
	and eat to fuel our brains
	every day.
	To reconcile this
	discrepancy between theoretical possibility and familiar reality 
	requires a 
	nonequilibrium thermodynamics of realistic computation: where nonequilibrium
	distributions---corresponding to metastable memories---are transformed 
	under practical constraints by
	controlled driving in finite time.
	From the perspective of nonequilibrium thermodynamics employed here,
	logical irreversibility is indeed compatible with thermodynamic reversibility if accompanied by
	a metastable increase in
	nonequilibrium-addition
	to free energy
	which can later be leveraged to reclaim the original work input.
	Hence: computation without dissipation.
	However, 
	our demands for speed and modularity each imply trade-offs that necessitate dissipation,
	while practical limitations of the controller's knowledge and dexterity further challenge the attainable thermodynamic efficiency of computation.
	Here we will develop a few of the 
	fundamental
	thermodynamic consequences of transforming metastable memories and identify
	several practical 
	opportunities for greater energetic efficiency.

	The following contains several new results, including:
	(1) A new decomposition of the nonequilibrium free energy that shows under what
	circumstances a coarse-grained description is sufficient to understand the
	thermodynamics of metastable memory transformations;
	(2) Implications for composite memory systems and 
	the role of knowledge in work extraction;
	(3) The thermodynamic cost of modular computation, which generalizes a recent
	result by Boyd et al.~\autocite{Boyd17}; and
	(4) The minimal work expected of any two-input--one-output logic function, and
	the dissipation incurred when these circuits are not designed for the statistics
	of the memories they transform.
	We close with a short tutorial that explicitly calculates the fundamental
	thermodynamic limits of the universal NAND gate.

	\section{Metastable memory systems}

	We start by considering a memory system, which is simply a physical system meant
	to store information.
	During computations,
	the dynamics of the memory system is driven by an external work reservoir
	to transform the memory from its initial state to its final state.
	For the memory system to be of much practical utility,
	it should be able to store memories robustly between computations.  
	One way to achieve this is with non-volatile memory elements that---through
	metastability---retain their memories over long timescales without active power
	consumption, even when the computer is turned off.\footnote{Two types of memory
		are common in practical computers.
		The first uses a non-volatile metastable memory that does not require energetic
		upkeep and retains its memory even when the computer is powered off.
		The second type requires active power to retain the memory,
		as in CMOS transistor architectures, 
		where the inevitable leakage of currents implies constant power consumption. 
		Without power, the volatile memory is lost.
		For reasons of both anticipated supremacy in energetic efficiency and clarity of
		our exposition,
		we choose to describe transformations of the former non-volatile type of memory
		in the following.  However, 
		we expect our results to maintain at least some relevance in the energetic
		limits of 
		transformations of active memories, where the work and dissipation discussed
		here should be roughly additive to the 
		background `housekeeping' power consumption by active circuits.
	}  
	
	At each moment, the work reservoir can exert influences according to the vector
	quantities $x \in \chi$.  For example, $x$ may represent the
	configuration of the applied electromagnetic field, a collection of piston
	positions, or any other controllable factors that 
	influence the Hamiltonian of the memory system.
	The instantaneous Hamiltonian $\mathcal{H}_x$ 
	of the memory system 
	determines the instantaneous energies $\{ E_x(s) \}_{s \in \SSet}$ of the
	system's microstates $\SSet$.
	The control parameter
	$x$ is held fixed while the memory is to be retained.
	Changes to the memory system are implemented by  
	a trajectory of time-varying control $\drive$ (often called a `protocol' in the literature) over a duration $\tau$ that drives the system to a new state.
	
	Computations utilizing metastable memories imply a strong separation of
	timescales in the non-driven dynamics of the memory system, such that the
	dynamics of the various metastable regions are nearly autonomous with respect to
	each other
	and can quickly 
	establish local equilibria.
	The autonomy within certain regions of state space suggests that we partition
	the set of microstates $\SSet$ into a set of metastable \emph{memory states}
	$\MSet$.															
	
	The system is also in contact with an effectively-memoryless heat bath at
	temperature $T$ with which it exchanges energy, which enables the system's
	relaxation to both local and global equilibrium.

	Over very long times (times much longer than any computation performed by the
	system, and much longer even than the waiting times between computations), the
	memory system---if left undriven, experiencing only the static control setting
	$x$---would asymptotically relax to the \emph{global equilibrium distribution}
	$\GlobalEq[x]$, which is 
	exactly stationary under the combined influence of the Hamiltonian
	$\mathcal{H}_x$ and the interaction with the heat bath.
	We assume that the global equilibrium distribution is the canonical one:
	$\GlobalEq[x](s) = \frac{e^{-\beta E_{x}(s)}}{Z_x}$, where $\beta = (\kB
	T)^{-1}$ and $Z_x$ is the standard partition function $Z_x = \sum_{s \in \SSet}
	e^{-\beta E_x(s)}$ that normalizes the distribution and yields the equilibrium
	free energy $F_x^\text{eq} = - \kB T \ln Z_x$.
	
	However, 
	metastability implies that 
	this timescale of global relaxation is 
	much longer than the timescale of computation.
	On the timescale \emph{between} computations,
	all probability density within each memory state $m \in \MSet$ is assumed to relax
	approximately to its \emph{local-equilibrium distribution} $\LocalEq[x]{m}$,
	as discussed in \autocite{Espo12} for the case of strong separation of timescales,
	with:
	\begin{align}
	\LocalEq[x]{m}(s) 
	&= \delta_{s \in m} \, \frac{e^{-\beta E_{x}(s)}}{\sum_{s' \in m} e^{-\beta
			E_{x}(s')} } 	= \delta_{s \in m} \, \GlobalEq[x](s) \, \frac{Z_x}{\LocalZ[x]{m}} 	= \delta_{s \in m} \, e^{-\beta \left( E_x(s) - \LocalFEq[x]{m} \right) } ~,
	\end{align}
	where $\LocalZ[x]{m}$ is the memory's  \emph{local partition function}:
	$\LocalZ[x]{m} \equiv \sum_{s \in m} e^{-\beta E_{x}(s)} $.
	This quantity strongly suggests defining the 
	\emph{local-equilibrium free energy}:
	\begin{align}
	\LocalFEq[x]{m} \equiv - \kB T \ln \LocalZ[x]{m} ~,
	\end{align}
	which turns out to provide useful intuition for the thermodynamics of
	transformations between metastable states, as we shall soon see.

	Between computations,
	the distribution relaxes quickly to a classical superposition of local
	equilibria determined by the net probability in each memory state at the end of
	the last computation.  
	Given a post-computation distribution over memory states $\Pr(\MSt_\tau)$, the
	distribution over microstates quickly approaches the metastable superposition:
	\begin{align}
	\Pr(\St_{\tau + \delta t}) \approx \sum_{m \in \MSet} \Pr(\MSt_\tau = m) \,
	\LocalEq[x_\tau]{m} ~,
	\label{eq:MetastableDistr}
	\end{align}
	where $\St_t$ is the random variable for the microstate at time $t$
	and $\MSt_t$ is the random variable for the memory state at time $t$.

	However, during a computation, the dynamic control protocol $\drive$ induces a
	net state-to-state stochastic transition dynamic $\Transition$ over $\SSet$ that can strongly couple and
	transform memory states, as required of a computation.

	\section{Driven dynamics and computations}

\begin{figure}[h]
		\includegraphics[width=0.75\textwidth]{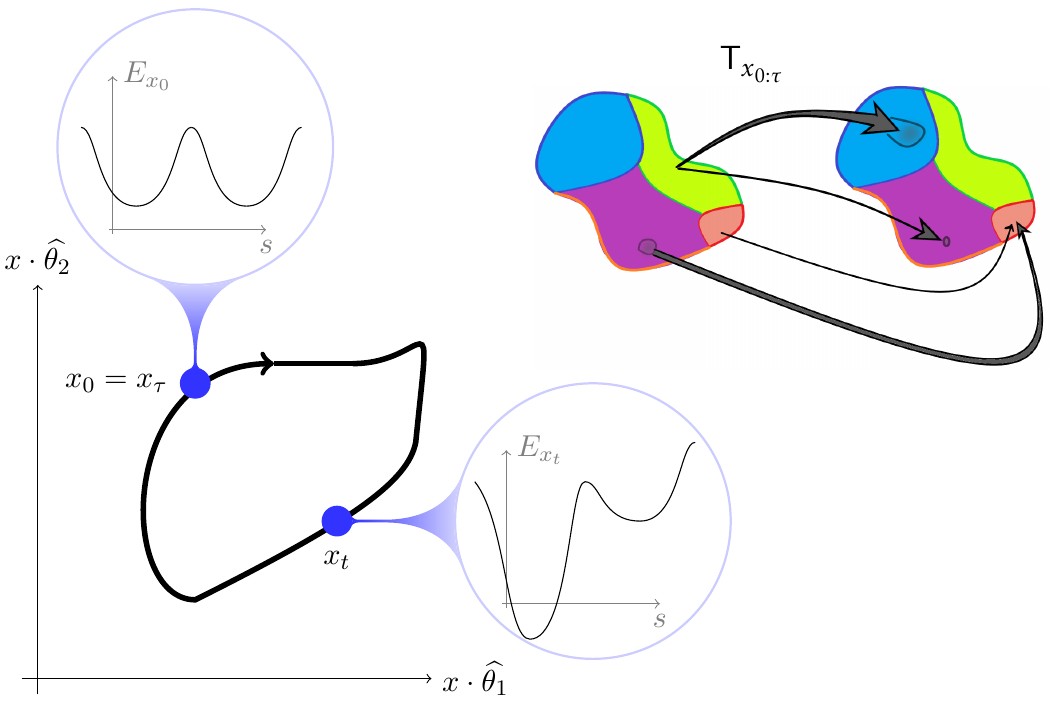}
		\caption{A generic driving protocol $\drive$ (the time series of the control
			parameters over a duration $\tau$) induces a net stochastic transition dynamic
			$\Transition$ over the set of microstates $\SSet$.  
			A thermodynamically efficient computation on the coarse-grained set of memory
			states $\MSet$ is implemented via a control protocol chosen from
			$\boldsymbol{\chi}_\mathcal{C}$ that reliably implements the logical
			transformation with minimal dissipation.}
			\label{fig:Driving4Comp}
	\end{figure}

	A deterministic \emph{computation} $\mathcal{C}: \MSet \to \MSet$ is an
	operation mapping the set of memory states $\MSet$ to itself. In practice, it is
	implemented by a driving protocol $x_{0:\tau}$ that controls the evolution of
	the system for a duration $\tau$.  
	Fig.~1 (Left) shows schematically how the protocol can change the energy landscape 
	to induce (Right) a net transition among microstates that corresponds to a (generically stochastic) computation on the coarse-grained memory states (depicted as the different shaded regions).
	The system should start and end with the same resting influence
	$x_0 = x_\tau$, as can be seen in the top-left bubble of Fig.~\ref{fig:Driving4Comp}, if a consistent metastable memory landscape is desired between computations.
		The set of all protocols that \emph{reliably} implement a computation
	$\mathcal{C}$ is:
	\begin{align}
	\boldsymbol{\chi}_\mathcal{C} \equiv \Bigl\{ x_{0:\tau} \in \chi^{[0, \tau]} :
	\, \Pr_{\drive} \bigl( \St_\tau \notin \mathcal{C}(m) \big| \St_0 \sim
	\LocalEq[x_0]{m} \bigr)  < \epsilon \, \text{ for all } m \in \MSet \Bigr\} ~,
	\end{align}
	for some error tolerance $\epsilon$.
	The assumed separation of timescales allows us to employ the local equilibrium
	distribution $\LocalEq[x_0]{m}$ as the initial distribution in the test for
	reliable memory evolution.

	Different protocols implementing the same computation typically dissipate
	different amounts of work.
	The grand challenge for energy-efficient computation
	is to identify the 
	control protocols, given realistic control restrictions,
	that reliably implement a computation in finite time with the minimal resultant
	work dissipation.

	\section{Work, nonequilibrium free energy, and dissipation}
	
	The work dissipated in a computation is the work $W$ that is irretrievably lost
	to the environment:
	\begin{align}
	\Wdiss = W - \Delta E_{x_t}(s_t) - \kB T \Delta \ln \bigl(
	\Pr_{\drivinghistory}(s_t) \bigr) ~,
	\label{eq:diss}
	\end{align} 
	where the right-most term is recognized as the change in the non-averaged
	precursor to nonequilibrium entropy.
	Equivalently, $\Wdiss$ is the amount of heat 
	that is \emph{not} offset by a corresponding change in this trial-specific
	internal entropy.
	Notably, Eq.~\eqref{eq:diss} is valid over \emph{any} time interval
	$t \in [0, \tau]$, and therefore (by adjusting the considered duration $\tau$)
	also tracks the dynamics of the dissipation associated with the computation as
	the system relaxes during and after the work performed to implement the
	computation.
	As the memory system relaxes to a new metastable distribution on a
	relatively short timescale relevant to computation, the dissipation should
	appear to saturate to the total dissipation associated with the computation.  
	Above,  $\Pr_{\drivinghistory[t]}(s_t)$ is the \emph{driving-induced}
	probability of the microstate $s_t$ at time $t$.
	Similarly, $\Pr_{\drivinghistory[t]}(\St_t)$ is the driving-induced probability
	\emph{distribution} over microstates---it can be thought of as the distribution
	over microstates \emph{conditioned} on the full driving history
	$\drivinghistory[t]$, including both the controlled preparation of the system
	prior to the computation and the protocol implementing the computation, up to
	time $t$.

	The \emph{expected} dissipation, given some initial preparation of the memory
	system and a particular driving protocol, is thus:
	\begin{align}
	\braket{\Wdiss} = \braket{W} - \Delta \mathcal{F} ~,
	\label{eq:originalWdiss}
	\end{align} 
	where the 
	expected nonequilibrium free energy at time $t$ is:
	\begin{align}
	\mathcal{F} 
	&= U - \kB T \, \Shannon \bigl( \Pr_{\drivinghistory[t]}(\St_t) \bigr) 
	= F_{x_t}^\text{eq} + \kB T \, \DKL \bigl( \Pr_{\drivinghistory[t]} (\St_t)
	\big\| \GlobalEq[x_t]  \bigr) ~.
	\end{align}
	Above, $U = \braket{ E_{x_t}(s_t) }_{\Pr_{\drivinghistory} (\St_t)}$ is the
	expected internal energy of the system at time $t$.
	$\Shannon ( \cdot )$ is the Shannon entropy of its  argument, and we will use
	$\Shannon_{\drivinghistory[t]} (\St_t) $ to denote
	$\Shannon \bigl( \Pr_{\drivinghistory[t]}(\St_t) \bigr)$.  
	Finally, $\DKL( \cdot )$ is the Kullback--Leibler divergence,
	which is always non-negative.
	$\DKL \bigl( \Pr_{\drivinghistory[t]} (\St_t) \big\| \GlobalEq[x_t]  \bigr) $ is
	the \emph{nonequilibrium addition} to free energy---the thermodynamic resource
	corresponding to the distribution being out of equilibrium. 
	It should be noted that $\Delta F_{x_t}^{\text{eq}} = 0$ over the full course of
	a computation since a computation starts and ends with the same resting
	influence $x_0 = x_\tau$.
	
	Recent finite-time fluctuation theorems (most directly: Eqs.~(38) and (42)
	of Ref.~\autocite{Riec17}) guarantee that the average dissipated work, when starting
	in any (potentially non-equilibrium and non-steady-state) distribution, is
	always non-negative: 
	\begin{align}
	\braket{\Wdiss} \geq 0 ~.
	\end{align}
	This implies that the average amount of work that must be performed is bounded
	by:
	\begin{align}
	\braket{W} \geq \Delta \mathcal{F}~.
	\label{eq:WgeqDF}
	\end{align}
	Eq.~\eqref{eq:WgeqDF} can also be derived by other means,
	as in Refs.~\autocite{Taka10, Espo11}. 
	The work performed in surplus to $\Delta \mathcal{F}$ is eventually dissipated
	and contributes to the entropy production by the computation.

	So, how much work is actually dissipated?
	Surely, the average work dissipated in transforming a distribution depends on the particulars of the protocol, with plenty of room for wastefulness if the protocol is not carefully designed.  However, the \emph{minimal} dissipated work is characterized by the allowed duration $\tau$ to implement the transformation and also by the degree of control one has over the system's Hamiltonian.  Let us briefly consider the case of perfect control, in which the controller can apply any Hamiltonian to the system.
	By instantaneously changing the initial Hamiltonian---to make any initial distribution the canonical distribution of the new Hamiltonian---before subsequent finite-speed driving of the system, we can immediately apply the recent results 
	of finite-time thermodynamics~\autocite{Sala83, Siva12,
		Zulk14, Bona14, Mand16} 
	to conclude
	that the work dissipated by an optimal
	protocol---meant to transform between two distributions in a finite time $\tau$
	with minimal dissipation---is
	generically (to first order of approximation) inversely proportional to the
	allowed duration $\tau$.  I.e.: $\braket{\Wdiss}_\text{min} \sim \tau^{-1}$.
	In part, the next section will show that this same $\tau^{-1}$ scaling of the dissipation can be achieved at intermediate timescales as long as the distribution stays close to a local-equilibrium distribution, even if it is never close to a global equilibrium distribution.
	More generally, the next section contains the fundamental thermodynamics of computation through the transformation of metastable memories.

	\section{Nonequilibrium thermodynamics, at level of memory states}

	The above is all now-standard nonequilibrium thermodynamics.
	However, we seek thermodynamic implications for transformation of \emph{memory
		states} rather than microstates.
	Fortunately, a rigorous hierarchical description
	can be achieved through a series of decompositions of familiar thermodynamic
	quantities.

	To start, we note that since $\MSet$ is a coarse-graining of $\SSet$, we have: 
	\begin{align}
	\Shannon_{\drivinghistory[t]} (\St_t ) 
	&= \Shannon_{\drivinghistory[t]} ( \MSt_t, \St_t ) 		= \Shannon_{\drivinghistory[t]} (\MSt_t ) + \Shannon_{\drivinghistory[t]} (\St_t
	| \MSt_t ) ~.
	\end{align}
	We will use the above together with a novel decomposition of the expected
	internal energy, which is valid at any time given any coarse-graining of
	microstates $\SSet$ into the coarse-grained set $\MSet$.  In particular, the
	expected internal energy of the system can be decomposed as: 
	\begin{align}
	U 
	= \braket{ E_{x_t}(s_t) }_{\Pr_{\drivinghistory}(\St_t)} 
	= 
	\bigl\langle \braket{ E_{x_t}(s_t) }_{\Pr_{\drivinghistory} (\St_t| \MSt_t = m)}
	\bigr\rangle_{ \Pr_{\drivinghistory} (\MSt_t = m) } 
	~.
	\end{align}
	Crutially, utilizing the identity
	$E_x(s) = -\kB T \ln \bigl( \LocalEq[x]{m}(s) \bigr) + \LocalFEq[x]{m}$, we find
	that the expected internal energy, if the system has been driven into memory
	state $m$ is:
	\begin{align}
	\braket{ E_{x_t}(s_t) }_{\Pr_{\drivinghistory} (\St_t| \MSt_t = m)} = \kB T \,
	\Shannon_{\drivinghistory}( \St_t| \MSt_t = m ) + \LocalFEq[x_t]{m} +
	\LocalFadd[t]{m}~.
	\end{align}
	Above, $\LocalFEq[x_t]{m}$ is the local-equilibrium free energy and 
	\begin{align}
	\LocalFadd[t]{m} \equiv \kB T \DKL \bigl( \Pr_{\drivinghistory}(\St_t| \MSt_t =
	m) \big\| \LocalEq[x_t]{m} \bigr) 
	\end{align} 
	is the \emph{local
		nonequilibrium addition to free energy} in region $m$.
	The expected internal energy thus always has the decomposition:
	$U = \kB T \Shannon_{\drivinghistory} (\St_t | \MSt_t ) +
	\braket{\LocalFEq[x]{m}}_{\Pr_{\drivinghistory}(\MSt_t)} + \braket{
		\LocalFadd[t]{m}  }_{\Pr_{\drivinghistory}(\MSt_t)}$.
	At the same time, we always have that $U =  \mathcal{F} + \kB T
	\Shannon_{\drivinghistory} (\St_t)$.
	Putting these together, we find that the nonequilibrium free energy can always
	be decomposed according to the contributions commensurate with the
	coarse-grained description:
	\begin{align}
	\mathcal{F} = \braket{\LocalFEq[x_t]{m}}_{\Pr_{\drivinghistory}(\MSt_t)} +
	\braket{ \LocalFadd[t]{m} }_{\Pr_{\drivinghistory}(\MSt_t)}
	- \kB T \Shannon_{\drivinghistory} ( \MSt_t ) ~.
	\label{eq:DecomposedF}
	\end{align}
	
	Moreover, when the coarse graining is according to well-designed metastable
	memory states, the separation of timescales implies that \\$\LocalFadd[t]{m} \to
	0$ quickly after any driving.\footnote{It is important to note that this is an
		\emph{assumption} about the dynamics which is well-suited to the memory systems
		typically used in practical computations.  The results of the following are only
		as reliable as this assumption is valid.}  Hence, before and shortly after a
	computation, we can decompose the nonequilibrium entropy into two very
	manageable parts:
	\begin{align}
	\mathcal{F} \approx 
	\braket{\LocalFEq[x_t]{m}}_{\Pr_{\drivinghistory}(\MSt_t)}
	- \kB T \Shannon_{\drivinghistory} ( \MSt_t ) ~;
	\label{eq:ParrDecomp}
	\end{align}
	that is:
	the expected local-equilibrium free energy,
	less the coarse-grained entropy of the memory states.
	Any difference from equality is due to work already performed that is expected
	to soon be dissipated in the relaxation to local equilibria. 
	Eq.~\eqref{eq:ParrDecomp} was previously highlighted in \autocite{Parr15}. 
	However the local nonequilibrium addition to the free energy, as in Eq.~\eqref{eq:DecomposedF},
	is a new finding that offers further insight.
	
	The local nonequilibrium addition to free energy is a thermodynamic resource that in principle can be traded to perform useful work.  However, if either (1) the timescale of relaxation within each memory state is faster than the relevant speed of the driving protocol or (2) the control parameters are too coarse or otherwise incapable of influencing the fine degrees of freedom within the memory state, then local nonequilibrium addition to free energy will inevitably be lost to dissipation as the local distributions relax to their local equilibria.
	Conversely, the coarse-grained memory probabilities are assumed to be metastable and controllable, and so nonequilibrium changes at the coarse-grained level can be implemented thermodynamically reversibly.

	With these considerations in mind,
	Eq.~\eqref{eq:DecomposedF} suggests that a driving protocol that keeps the
	distribution close to a metastable one (i.e., a weighted average of
	local-equilibrium distributions) at all times, such that  $\LocalFadd[t]{m}$
	always stays close to zero in each metastable region,
	can be used to implement thermodynamically-efficient computations with
	$\braket{\Wdiss} \sim \tau^{-1} \to 0$ as $\tau \to \infty$.
	In this nearly-quasistatic limit, such processes will be thermodynamically reversible.
		Finally, coming back to Eq.~\eqref{eq:originalWdiss},
	this implies that the minimum average work necessary to implement a computation
	on metastable memory states is:
	\begin{align}
	\braket{W}_\text{min} = \Delta
	\braket{\LocalFEq[x_t]{m}}_{\Pr_{\drivinghistory}(\MSt_t)}
	-  \kB T \Delta \Shannon_{\drivinghistory} ( \MSt_t ) ~.
	\label{eq:GeneralizedLandauer}
	\end{align}
	
\begin{figure}[h]
		\includegraphics[width=0.5\textwidth]{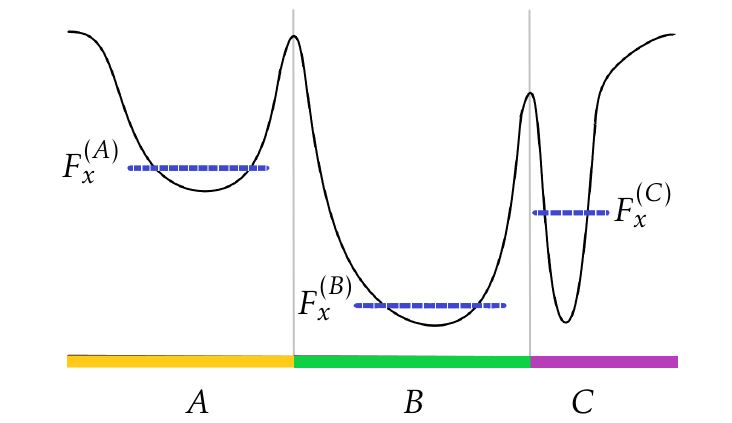}
		\caption{Schematic of a simple one-dimensional potential with three metastable
			wells of different character.  Metastability requires barriers of at least
			several times the thermal energy $\kB T$.  
			Larger local-equilibrium free energy $F_x^{(m)}$ can result from higher energy
			or greater certainty---or both.}
		\label{fig:EnergyLansdscape}
	\end{figure}

	In the computational setting, Eq.~\eqref{eq:GeneralizedLandauer} can be interpreted as a generalization of Landauer's
	principle for the minimum work necessary to implement computations that
	transform metastable memories of \emph{different} local-equilibrium free energies.
	In a more general setting, 
	when $\braket{W}_\text{min}$ is negative, $\braket{W_\text{extracted}}_\text{max} = -
	\braket{W}_\text{min}$ can also be interpreted as the 
	maximal work that can be \emph{extracted} from a metastable system---and the heterogeneity of local free energy then offers an easy explanation of how a single bit of macroscopic information (e.g., "Is the apple in the left or right box?") can carry a macroscopically huge (much larger than $\kB T \ln 2$) energetic gain, as in \autocite{Gokl17}, when interacting with far-from-equilibrium systems.
	Fig.~\ref{fig:EnergyLansdscape} gives further intuition for the meaning of the local-equilibrium free energies in Eq.~\eqref{eq:GeneralizedLandauer}: larger local-equilibrium free energy can result from either larger average energy or from more certainty in the local-equilibrium microstate distribution.

	Clearly, if all local-equilibrium free energies are equal (either through
	identically-constructed potentials or otherwise through some delicate balance of
	local energies and entropies), such that $\LocalFEq[x_0]{m} =
	\LocalFEq[x_0]{m'}$ for all $m, \, m' \in \MSet$, then $\Delta
	\braket{\LocalFEq[x_t]{m}}_{\Pr_{\drivinghistory}(\MSt_t)} = 0$ and the minimal
	work necessary to implement a computation reduces to:
	\begin{align}
	\braket{W}_\text{min} 
	&= 
	-  \kB T \Delta \Shannon_{\drivinghistory} ( \MSt_t )
			=  \kB T  \Shannon_{\drivinghistory[0]} ( \MSt_0 ) -  \kB T 
	\Shannon_{\drivinghistory[\tau]} ( \MSt_\tau )
	~.
	\label{eq:Landauer}
	\end{align}
	such that the minimum work depends solely on the change in entropy over memory
	states.
	Eq.~\eqref{eq:Landauer} constitutes the modern understanding of Landauer's
	principle for the minimum work necessary to implement computations that
	transform metastable memories of equal free energy.
	
	If the computation reduces the internal entropy of the memory system, then it
	will require work (although this work can later be reclaimed and recycled if the
	computation was performed without dissipation).
	Conversely, when $\braket{W}_\text{min}$ is negative,
	work can be \emph{extracted} as a result of the transformation, for example to
	lift a weight or to fuel other computations.  In this latter case, the memory
	system can perform as an \emph{information engine}, trading certainty for useful
	work.
	In this regime, $\braket{W_\text{extracted}}_\text{max} = -
	\braket{W}_\text{min} =  \kB T \Delta \Shannon_{\drivinghistory} ( \MSt_t )$ is
	the \emph{maximal} average work that can be extracted from transformations that
	result in this memory-entropy change.

	\section{Implications for composite memory systems}

	If the system is composed of $N$ multi-stable memory elements, then it is
	natural to treat the memory state as a composite state of $N$ random variables:
	$\MSt_t = \MSt_t^{(1:N+1)} = \bigl( \MSt_t^{(1)} , \MSt_t^{(2)} , \dots
	\MSt_t^{(N)} \bigr)$.
	In general, the memory elements are correlated, such that
	$\Pr_{\drivinghistory[t]} \bigl( \MSt_t^{(1)} , \MSt_t^{(2)} , \dots
	\MSt_t^{(N)} \bigr) \neq \prod_{n=1}^{N} \Pr_{\drivinghistory[t]} \bigl(
	\MSt_t^{(n)} \bigr) $.
	Moreover, this correlation has important thermodynamic consequences.

	Consider a nonequilibrium system of $N$ identical memory elements, each having
	$K$ identical metastable regions 
	(such that $\LocalFEq[x_0]{m} = \LocalFEq[x_0]{m'}$ for all $m, \, m' \in
	\MSet$),
	and 
	let $h_{\drivinghistory[t]} \equiv \frac{\Shannon_{\drivinghistory[t]} (
		\MSt_t^{(1:N+1)} )}{N}$ denote the coarse-grained entropy \emph{density} of the
	memory system.
	If the system is transformed by some driving protocol $\drive$ that increases
	the system's entropy,
	then, from Eq.~\eqref{eq:Landauer}, the maximal work (per memory element) that
	could have been extracted in the process is given by:
	\begin{align}
	\tfrac{\beta}{N} \braket{W_\text{extracted}}_\text{max} =
	h_{\drivinghistory[\tau]} - h_{\drivinghistory[0]} ~.
	\label{eq:EntropyDensity}
	\end{align}

	In the case where the memory elements are arranged in a 1-dimensional topology,
	the entropy density has been taken to mean the Shannon entropy \emph{rate} of
	the sequence as it is scanned spatially~\autocite{Boyd17JSP}.
	While this is technically correct, it is sufficiently nuanced to require careful
	interpretation.
	In particular, the entropy density $h \equiv \lim_{L \to \infty}
	\tfrac{\Shannon( \boldsymbol{p}_L ) }{L}$ that can be inferred from the
	frequentist statistics gathered along the sequence of the instantaneous
	configuration will in general converge to a value that is \emph{not} 
	the same as $h_{\drivinghistory}$.\footnote{For example, 
		the string of the digits of pi ($3.14159 \dots$) has a Shannon entropy rate of 
		$h = \log_2(K)$ bits-per-symbol in any $K$-ary expansion for $K \in \{ 2, 3,
		\dots \}$. (E.g., $h = \log_2(10)$ in the given decimal expansion of pi, whereas
		$h = 1$ in its binary expansion.)
		It would naively seem to provide no thermodynamic fuel as it appears to be a
		completely `random' sequence.  To the contrary, if stored into memory, the
		sequence contains \emph{full} thermodynamic fuel (i.e., it can be fully
		leveraged to do work)
		because the memory system will be \emph{driven} by $\drivinghistory$ to uniquely
		hold this sequence, so that $\Shannon_{\drivinghistory}(\MSt_t) = 0$ since
		$\Pr(\MSt_t) = \delta_{\MSt^{(1)}, 3} \delta_{\MSt^{(2)},  1}
		\delta_{\MSt^{(3)}, 4} \dots \,$.  I.e., in the space \emph{of sequences}, the
		state of the memory system is delta-distributed. 
		This particular example 
		points to the proper way to think about entropy and about what \emph{kind} of
		information can be thermodynamically leveraged in computer memory in general.
	}
	Rather:
	\begin{align}
	\ln(K) \geq h \geq h_{\drivinghistory[{}]} \geq 0 ~.
	\end{align}
	Crucially, it is the entropy \emph{conditioned on the driving history} that
	matters in the theoretical limit of
	what orderliness can be thermodynamically leveraged---and this is a priori
	distinct from anything that could be inferred from the instantaneous
	configuration, even in the limit of $N \to \infty$.

	Moreover, this thermodynamic entropy density is independent of spatial
	dimension, or even any spatial topology of the memory elements.
	Indeed, the topology of the memory elements---being arranged in a 1-dimensional
	string for example---is a priori \emph{independent} of the topology of the
	correlations among random variables.  And it is only the latter that
	fundamentally matters for the thermodynamics of information processing.
	However, spatial locality occasionally does correspond to the correlational
	structure, especially when correlations develop as a consequence of local
	physical interactions.

	Turing machines and related models of computation require not only a bit string
	but also a memoryful read--write head that can operate on the tape.  Treating
	these two components inclusively as part of the memory system makes the system
	self-contained and provides important lessons about the thermodynamics of such
	functionally segregated systems~\autocite{Boyd17PRL, Boyd17JSP}.

	This sort of inclusiveness\footnote{ 
		Including two subsystems (say, `subsystem' and `demon') explicitly as components of the same memory system means that the coarse-grained memory entropy decomposes as:
		$\Shannon ( \MSt^{(\text{sub})} ,\MSt^{(\text{demon})} )  = \Shannon (
		\MSt^{(\text{sub})} ) + \Shannon ( \MSt^{(\text{demon})} ) - \MI (
		\MSt^{(\text{sub})} ;\MSt^{(\text{demon})} )$,
		where $\MI$ denotes mutual information.  Interpreting the latter quantity as the knowledge the demon has of the system, we can see from substitution into Eq.~\eqref{eq:GeneralizedLandauer} that 
		knowledge is another thermodynamic resource that can be exchanged for entropy reduction, free-energy gain, or work extraction; but we also see that its origination carries either an energetic or entropic cost of at least what the knowledge was later worth---knowledge is a \emph{medium} for thermodynamic transactions rather than a free source of energy.  
	}
	also sheds light on `Maxwell's demon'-type scenarios 
	where the net system is decomposed into a subsystem and a `demon'.
	If the subsystem is initially out of equilibrium, then the demon can simply
	extract work by extracting the subsystem's nonequilibrium addition to free
	energy---no mystery there.  However the more complicated story arises when the
	subsystem is initially \emph{in} equilibrium, as in Maxwell's original
	gedanken-experiment where the subsystem is a two-compartment box of gas starting
	in equilibrium.
	The demon can nevertheless decrease the entropy of the subsystem by increasing
	its knowledge of the subsystem~\autocite{Lloy89}.
	Work can then be extracted as the subsystem is subsequently brought back to
	equilibrium.
	But if this process is to reset to form a full cycle---if the demon's memory is
	to be erased for example---then no net work can be extracted on average. 
	Interestingly, it is not \emph{necessarily} erasure where cost is
	incurred~\autocite{Boyd16}, but whether the cost is incurred in measurement or
	in erasure can be explained via heterogeneous local-equilibrium free energies of
	the memory states as an application of Eq.~\eqref{eq:GeneralizedLandauer}.

	To close this topic, we note that work \emph{can} be extracted at a constant
	rate when 
	an active environment continuously drives the subsystem out of equilibrium, at
	no cost to the extractor (formerly known as `demon').  The extractor
	can then siphon off the power that it needs to sustain its luxurious
	nonequilibrium lifestyle---perhaps even to appease its greed for massive speedy
	computations.

	\section{The thermodynamic cost of ignorance and neglect}

	If one has sufficient control over the energy levels of a system, and if
	quasistatically-slow transformations are 
	tolerable, then a transformation between any two distributions is always
	possible \emph{without dissipation}.
	Example methods to implement such dissipationless computations are given, for
	example, in \autocite{Aabe13, Garn17, Parr15, Boyd17}.
	We have further argued that the zero-dissipation limit is also approached with the slightly weaker 
	requirement that 
	metastable distributions (rather than strictly the global equilibrium distribution) are maintained throughout the transformation.
	However, even in this nearly-quasistatic case, if correlations are ignored (or if 
	other features of the distribution are neglected or mis-represented for whatever
	reason in the manipulation of the distribution),
	then there is necessarily extra work incurred and dissipated when the driving
	protocol is run.

	Suppose a driving protocol $\drive^*$ is chosen to minimize dissipation while
	implementing a computation $\mathcal{C}$ and starting in the distribution
	$\asif$.
	I.e.,
	$\drive^* \equiv \argmin_{\drive \in \boldsymbol{\chi}_{\mathcal{C}}}
	\braket{\Wdiss}_{\Pr_{\drive}(\St_{0:\tau} | \St_0 \sim \asif)}$.
	For simplicity, let's further assume that we take the $\tau \to \infty$ limit so
	that this computation is achieved with no dissipation at all when the system is
	initiated in the distribution $\asif$.

	Moreover, $\drive^*$ and $\asif$ are a pair,
	each minimizing dissipation for the other.
	Indeed, 
	with $\braket{\Wdiss} = 0$ and the generic requirement that $\braket{\Wdiss}
	\geq 0$: it follows that, among possible ways to initiate the distribution over
	system states, 
	$\asif$ minimizes the dissipation incurred when running the drive protocol
	$\drive^*$.
	This latter fact allows us to utilize the recent theorem by Kolchinsky and
	Wolpert \autocite{Kolc17}: 
	that if $\asif$ minimizes the dissipation for some drive protocol $\drive$, then
	there is necessarily an extra dissipation incurred by starting in some other
	distribution $\actual$, given by:
	\begin{align}
	\beta \braket{\Wdiss(\actual)} - \beta \braket{\Wdiss(\asif)} =
	\DKL( \actual \| \asif ) -  \DKL( \actual[\tau] \| \asif[\tau] ) ~,
	\label{eq:KW_thm}
	\end{align}
	where $\braket{\Wdiss(\actual)} \equiv
	\braket{\Wdiss}_{\Pr_{\drive^*}(\St_{0:\tau} | \St_0 \sim \actual )}$,
	and 
	$\braket{\Wdiss(\asif)} \equiv \braket{\Wdiss}_{\Pr_{\drive^*}(\St_{0:\tau} |
		\St_0 \sim \asif )}$ which is equal to 0 in this case.
	Above, $\actual[\tau] = \Pr_{\drive^*}(\St_\tau | \St_0 \sim \actual)$ and
	$\asif[\tau] = \Pr_{\drive^*}(\St_\tau | \St_0 \sim \asif)$ 
	are the time-evolved versions of $\actual$ and $\asif$,
	respectively, under the influence of the driving $\drive^*$.

	There are several immediate novel consequences of Eq.~\eqref{eq:KW_thm} when
	applied to our framework, that are worth teasing out since they yield important
	general lessons about dissipation incurred during computation.

	\subsection{Dissipation through modularity and neglected correlation}

	Let us say that the system is \emph{actually} in distribution $\actual$, but the
	controller \emph{thinks}---or otherwise acts \emph{as if}---the distribution is
	$\asif$.

	One case in which this happens in practice is when correlations exist among
	parts of a memory system but computations are implemented only modularly. 
	Modular computing---by implicitly marginalizing over some of the memory
	elements---necessarily ignores the correlations among modular units.

	Suppose for example that we partition the memory system into two composite
	pieces 
	$\St_t = \bigl(\St_t^{(1)}, \St_t^{(2)} \bigr)$
	and that the two memory subsystems are correlated:
	$\actual[t] = \Pr(\St_t^{(1)}, \, \St_t^{(2)}) \neq \Pr(\St_t^{(1)}) \,
	\Pr(\St_t^{(2)})$;  
	but the two memory subsystems are operated on independently (i.e., modularly),
	which means 
	$ \asif[t] = \Pr(\St_t^{(1)}) \, \Pr(\St_t^{(2)})$. 
	I.e., the distribution, although correlated, is operated on \emph{as if} the two
	components were statistically independent.
	The implications are immediately accessible:
	\begin{align}
	\beta \ModularityDiss 
	&= \DKL( \actual \| \asif ) -  \DKL( \actual[\tau] \| \asif[\tau] ) 
	\\
	&= - \Delta \DKL \bigl( \Pr(\St_t^{(1)}, \, \St_t^{(2)}) \, \big\| \,
	\Pr(\St_t^{(1)}) \, \Pr(\St_t^{(2)}) \bigr) \\
	&= \MI \bigl( \St_0^{(1)} ; \, \St_0^{(2)} \bigr) - 
	\MI \bigl( \St_\tau^{(1)} ; \, \St_\tau^{(2)} \bigr) ~,
	\label{eq:MIchange}   
	\end{align}
	where $\MI( \cdot ; \cdot )$ is the mutual information between its arguments.
	This means that work is necessarily dissipated
	whenever a modular computation discards information between two
	subsystems.\footnote{The opposite situation (i.e., mutual information
		\emph{increasing} between the two subsystems) does not happen under the current
		assumption of modularity, and so we are not in danger of deriving
		$\braket{\Wdiss(\actual)} < 0$ here, which would be counter to the second law of
		thermodynamics.  If a computation creates correlation between two subsystems,
		then $\asif[\tau]$ would not be separable, and the analysis would have proceeded
		differently.}

	When the modular computations are being performed on metastable memory states,
	then assuming the memory starts and ends in a metastable distribution with
	microstate probabilities $\actual[t](s) = \sum_{m \in \MSet} \actual[t]'(m)
	\LocalEq[t]{m}(s)$ for $t = 0$ and $t=\tau$,
	we find that we can formulate the result in terms of the memory states of the
	two subsystems:
	\begin{align}
	\beta \ModularityDiss 
	&= \MI \bigl( \MSt_0^{(1)} ; \, \MSt_0^{(2)} \bigr) - 
	\MI \bigl( \MSt_\tau^{(1)} ; \, \MSt_\tau^{(2)} \bigr) ~.
	\label{eq:MIchangeforMSt}   
	\end{align}

	Although we have arrived at this result by rather different means,
	Eq.~\eqref{eq:MIchangeforMSt} is essentially the same as the main result of
	\autocite{Boyd17} (although there it was assumed that $\MSt^{(2)}$ is
	unchanged by the computation, which led to $\MSt_\tau^{(2)} = \MSt_0^{(2)}$
	there, and it was also assumed there that the local-equilibrium free energies
	were all the same).
	It is notable that our result does \emph{not} 
	require 
	any assumption about the local free energies of the memory subsystems---they can
	be arbitrarily heterogeneous.

	With modular computations happening on $N$ different subsystems, the result
	generalizes easily.
	With $\actual[t] = \Pr( \St_t^{(1)}, \, \St_t^{(2)} ,  \dots \,  \St_t^{(N)} )$
	and 
	$\asif[t] = \prod_{n=1}^N \Pr(\St_t^{(n)})$,
	we find that:
	\begin{align}
	\beta \ModularityDiss   
	&= - \Delta \DKL \Bigl( \Pr(\St_t^{(1)}, \, \St_t^{(2)} , \dots \,  \St_t^{(N)})
	\, \Big\| \, \prod_{n=1}^N \Pr(\St_t^{(n)}) \Bigr) \\
	&= \TC \bigl( \St_0^{(1)}, \, \St_0^{(2)} ,  \dots \,  \St_0^{(N)} \bigr) - 
	\TC \bigl( \St_\tau^{(1)}, \, \St_\tau^{(2)} ,  \dots \,  \St_\tau^{(N)} \bigr)
	~.
	\label{eq:TCchange}   
	\end{align}
	where $\TC  \bigl( \St_t^{(1)}, \, \St_t^{(2)} ,  \dots  \, \St_t^{(N)} \bigr) =
	\Bigl( \sum_{n=1}^N \Shannon ( \St_t^{(n)}) \Bigr) - \Shannon \bigl(
	\St_t^{(1)}, \, \St_t^{(2)} ,  \dots \,  \St_t^{(N)} \bigr)$ is the so-called
	\emph{total correlation} among its arguments.
	This generalization is necessary for predicting the
	dissipation that will be incurred when many modular computations are performed
	in parallel.

	Suppose instead we want to consider the problem at the level of metastable
	memory states, with the joint distribution over the memory states of the
	subsystems $\actual[t]' = \Pr( \MSt_t^{(1)}, \, \MSt_t^{(2)} ,  \dots \,
	\MSt_t^{(N)} )$.
	If the memory system is assumed to start and end the computation in a
	classically superposed metastable distribution such that
	$\actual[t] = \sum_{m \in \MSet} \actual[t]'(m) \LocalEq[t]{m}$
	at $t = 0, \tau$, as in Eq.~\eqref{eq:MetastableDistr}, then using 
	$\frac{\actual(s)}{\asif(s)} = \frac{\actual'(m(s))}{\asif'(m(s))}$ and making
	use of $\LocalEq[t]{m}(s) = \delta_{s \in m} \LocalEq[t]{m}(s)$ in our
	calculation,
	regardless of any heterogeneity among the local-equilibrium free energies,
	we again find that:
	\begin{align}
	\beta \ModularityDiss   
	&= \TC \bigl( \MSt_0^{(1)}, \, \MSt_0^{(2)} ,  \dots \,  \MSt_0^{(N)} \bigr) - 
	\TC \bigl( \MSt_\tau^{(1)}, \, \MSt_\tau^{(2)} ,  \dots  \, \MSt_\tau^{(N)}
	\bigr) ~.
	\label{eq:TCchangeforMSt}   
	\end{align}

	Whether framed in terms of microstates or memory states, 
	our general result means that: \emph{the minimal extra dissipation incurred by
		modular computation is exactly $\kB T$ times the reduction in total correlation
		among all memory subsystems.}

	\subsection{Dissipation through failing to model statistics of manipulated
		memory}

	Let us consider the implications for the common logic gates that serve as the
	building blocks for practical computers.  Recall that the simple NAND 
	gate is 
	sufficient for universal computation.  It is therefore worthwhile to consider
	what dissipation is commonly incurred in these important logic gates---and to
	show how this dissipation can be avoided.

	It is important to note that \emph{even without correlation} it is critical to
	correctly model the input statistics of a computation in order to avoid
	dissipating work.
	Modeling correlations is then a requirement on top of this.
	Since we have already briefly discussed the role of correlations, let us focus
	here on the more basic point of modeling input statistics 
	whatsoever.

	To address this, we will consider a physical instantiation of the memory
	components of a NAND gate, where we explicitly consider two memory elements
	whose memory states are to be used as the input for the NAND gate and another
	memory element that will store the value of the output.
	We will assume that only the output is over-written during the computation---the
	input memory states may be kept around for later use.

	Note that this is already a particular physical model of the NAND
	computation---indeed, alternatives exist such as storing the output in the
	location of one of the former inputs by over-writing one of the inputs.
	However, we will analyze the proposed two-input--one-output model here since it
	is arguably the most relevant to the typical desired use of a NAND gate.
	Other ancillary memory elements may be used in the computation as in
	\autocite{Owen17} but, since they will be returned to their original state
	by the end of the computation, these ancillary memories do not need to result in
	any additional dissipation and so can be left implicit in the 
	self-consistency of the
	current analysis.

	Each of the three explicitly-considered memory elements is assumed to be
	bistable (i.e., each of the three memory elements is assumed to have two
	metastable regions of state space).\footnote{For example, each memory element
		may be realized physically by the bistable magnetic moment of a
		superparamagnetic nanocrystal in the so-called `blocked' regime where the
		N\'{e}el relaxation time between metastable regions is much larger than the
		timescale of computation in the system.
		We can assume that
		there is sufficient uniaxial anisotropy (or sufficiently low temperature) to
		create a potential barrier many times the thermal energy between the potential
		wells of the two metastable regions.
		Although it is nice to have several realistic physical systems in mind,
		ultimately the physical details of the bistable memory element will be largely
		irrelevant, and the analysis transcends these specifics.}
	Let the microstate of each memory element be
	specified by its position in the interval $(-\pi, \pi]$.  
	Between computations, including at $t=0$ and $t=\tau$,
	the metastable regions for each memory element are $\zero \equiv (-\pi, 0]$ and
	$\one \equiv (0, \pi]$ which 
	gives a natural partition for the memory states.

\begin{figure}[h]
		\includegraphics[width=0.99\textwidth]{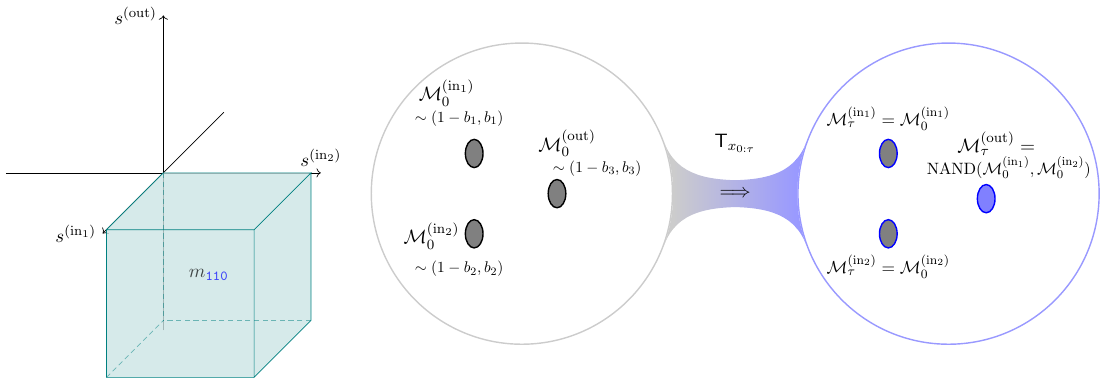} 
		\caption{Composite state space, memory states, and physical transformation
			associated with the logical NAND operation.}
		\label{fig:NAND}
	\end{figure}

	The microstate of the memory system at any time $t$ can be treated as a
	composite random variable $\St_t = (\St_t^{(\text{in}_1)},
	\St_t^{(\text{in}_2)}, \St_t^{(\text{out})})$ with $\St_t^{(\cdot)} \in (-\pi,
	\pi]$.
	Similarly, the memory state is the composite random variable  
	$\MSt_t =
	(\MSt_t^{(\text{in}_1)}, \MSt_t^{(\text{in}_2)}, \MSt_t^{(\text{out})})$ with
	$\MSt_t^{(\cdot)} \in \{ \zero, \one \}$ corresponding to the two metastable
	regions for each memory element.
	Thus, the joint state-space $\SSet = \mathbb{R}_{(-\pi, \pi]}^3$ has eight
	metastable regions,
	which we identify as the joint memory system's eight memory states:
	$\MSet = \{ m_{\zero \zero \zero}, m_{\zero \zero \one}, m_{\zero \one \zero},
	\dots m_{\one \one \one} \}$ where each memory state is labeled according to its
	corresponding region of state-space:
	$m_{j k \ell} = \bigl\{ s \in \SSet : \, s^{(\text{in}_1)} \in (0 - j \pi , \pi
	- j \pi], \, s^{(\text{in}_2)} \in (0 - k \pi , \pi - k \pi], \,
	s^{(\text{out})} \in (0 - \ell \pi , \pi - \ell \pi]  \bigr\}$.
	I.e., each of the memory states is one of the octants of state space, as shown in Fig.~\ref{fig:NAND} (Left).

	As a first analysis of this system, let us suppose that all memory elements are
	initially uncorrelated: 
	$\Pr(\MSt_0) = \Pr(\MSt_0^{(\text{in}_1)}) \Pr( \MSt_0^{(\text{in}_2)})
	\Pr(\MSt_0^{(\text{out})})$.
	Suppose though that each memory element has an initial bias such that
	$\Pr(\MSt_0^{(\text{in}_1)} = \one) = b_1$, $\Pr(\MSt_0^{(\text{in}_2)} = \one)
	= b_2$, and 
	$\Pr(\MSt_0^{(\text{out})} = \one) = b_3$.
	Let us call this distribution over memory states $\actual' = \Pr(\MSt_0)$.
	The corresponding initial distribution over microstates is:
	$\Pr(\St_0) = \sum_{m \in \MSet} \Pr(\MSt_0 = m) \, \delta_{\St_0 \in m} \,
	\LocalEq[x_0]{m}$, which we will call $\actual$.
	This physical setup, and the logical transformation of the output bit, is diagrammed in Fig.~\ref{fig:NAND} (Right).

	Suppose that a NAND gate is constructed such that it does not dissipate any work
	when it transforms the memory system in its equilibrium state: $\asif[0] =
	\sum_{m \in \MSet} \delta_{\St_0 \in m} \, \LocalEq[x_0]{m} / 8$.
	That is, the transformation is designed to dissipate no work when the memory
	states are all initialized in the uniform distribution $\asif' = \tfrac{1}{8} 
	\begin{bmatrix}
	1 & 1 & \dots & 1
	\end{bmatrix}$.

	The minimal extra dissipation incurred by using this NAND transformation
	(optimized for minimal dissipation in the case of uniform distribution over
	memory states), given that the initial statistics of the memory elements are
	actually biased by the $b_i$, is:
	\begin{align}
	\beta \MismatchDiss 
	&= \DKL( \actual \| \asif ) -  \DKL( \actual[\tau] \| \asif[\tau] ) \\ 
	&= \DKL( \actual' \| \asif' ) -  \DKL( \actual[\tau]' \| \asif[\tau]' ) \\ 
	&= \sum_{m \in \MSet} \actual'(m) \ln \left( \tfrac{\actual'(m)}{1/8} \right) -
	\actual[\tau]'(m) \ln \left( \tfrac{\actual[\tau]'(m)}{1/4} \right) \\
	&= \ln 8 - \Shannon (\actual') - \ln 4 + \Shannon (\actual[\tau]') \\
	&= \ln 2 - \boldsymbol{H}_2 (b_3) ~,
	\end{align}
	where $\boldsymbol{H}_2 (b) \equiv -b \ln b - (1-b) \ln (1-b)$.

	We reflect that the full dissipation of operating the NAND gate (when not
	optimized for the correct memory biases) is essentially the entropy production
	from ignoring the single bias of $b_3$ associated with the output.
	The protocol \emph{could} have been designed to be dissipation free, but the
	current NAND implementation does not erase $\MSt_0^{(\text{out})}$ in a way that
	salvages its original nonequilibrium addition to free energy.

	More generally, if we allow any kind of initial correlation among the initial
	configuration, such that 
	the input and output bits \emph{are} initially correlated according to
	$\actual' = \Pr(\MSt_0) \neq \Pr(\MSt_0^{(\text{in}_1)}) \Pr(
	\MSt_0^{(\text{in}_2)}) \Pr(\MSt_0^{(\text{out})})$,
	then the resulting dissipation generalizes to:
	\begin{align}
	\MismatchDiss = \kB T \ln 2 
	- \kB T \Shannon_{\actual'}\bigl( \MSt_0^{(\text{out})} \, \big| \,
	\MSt_0^{(\text{in}_1)} , \MSt_0^{(\text{in}_2)} \bigr) ~,
	\label{eq:Gen2to1dissipation}
	\end{align}
	where the right-most term is the conditional Shannon entropy of the
	\emph{initialized} value of the output bit (i.e., before the NAND operation is
	implemented), given the initial values of the input.

	Again, this $\beta \MismatchDiss$ turns out to be the irreversible entropy
	production of ignoring the nonequilibrium distribution of the original output
	bit.
	A smarter protocol could have instead leveraged this ordered nonequilibrium
	addition to free energy to perform the NAND computation with less work.
	But, since the protocol was not altered to take advantage of this initial
	nonequilibrium situation,
	the thermodynamic resource is forever lost through dissipation by the end of the
	computation.

	\subsection{Dissipation and minimal work for any two-input--one-output logic
		gate}

	Once having gone through this analysis, it should be clear that the NAND
	function played no essential role in determining the minimal dissipation from
	neglected initial biases, other than the fact that the NAND operation is a
	deterministic two-input--one-output function.
	Hence, Eq.~\eqref{eq:Gen2to1dissipation} describes the minimal dissipation of
	\emph{all} two-input--one-output logic gates---NAND, AND, XOR, etc.---when the
	output memory element is overwritten by a computation that does not leverage the
	initial biases of the memory elements it is manipulating.

	We focused above on the work \emph{dissipated} since this is the undesirable
	waste that designers of future hyer-efficient computers should be hyper-aware
	about.
	It is noteworthy though that even when no work is dissipated, the minimal
	\emph{work} to implement the computation will also depend on the initial biases
	of the memory elements that are to be manipulated.
	From Eq.~\eqref{eq:Landauer}, we see that the minimal work necessary to
	implement the NAND gate---and indeed to implement \emph{any}
	two-input--one-output gate where the output memory element is to be
	overwritten---is:
	\begin{align}
	\braket{W}_\text{min} 
	&= \kB T 
	\Shannon_{\actual'}\bigl( \MSt_0^{(\text{in}_1)} , \MSt_0^{(\text{in}_2)} ,
	\MSt_0^{(\text{out})} \bigr) - 
	\kB T 
	\Shannon_{\actual[\tau]'}\bigl( \MSt_0^{(\text{in}_1)} , \MSt_0^{(\text{in}_2)}
	, \MSt_\tau^{(\text{out})} \bigr) \\
	&= \kB T
	\Shannon_{\actual'}\bigl( \MSt_0^{(\text{out})} \big| \MSt_0^{(\text{in}_1)} ,
	\MSt_0^{(\text{in}_2)} \bigr) ~.
	\end{align}
	In the case of biased but uncorrelated initial inputs and outputs, this reduces
	to:
	$\braket{W}_\text{min} = \kB T \boldsymbol{H}_2 (b_3) $.
	However, as long as this work is not dissipated, it can continue to be salvaged
	and recycled in future computations.

	Looking back at $\braket{\Wdiss}$, we see that our result for the minimum work
	puts the dissipated work in a new context.
	In particular:
	\begin{align}
	\MismatchDiss = \kB T \ln 2 
	- \braket{W}_\text{min} ~.
	\label{eq:Gen2to1diss_vs_Wmin}
	\end{align}
	We interpret this as
	the minimum work that would need to be performed given the uniform distribution
	of initial memory states that the system was designed for, minus the minimum
	work given the actual biases.
	In the case of biased but uncorrelated initial memory states
	this can be framed as:
	\begin{align}
	\MismatchDiss = \kB T  \boldsymbol{H}_2 (\tfrac{1}{2})
	- \kB T  \boldsymbol{H}_2 ( b_3 ) ~.
	\end{align}
	When all memory states have the same local-equilibrium free energy, biases in
	the input should be treated as a resource---a nonequilibrium addition to free
	energy.  Ignoring these biases is to ignore and waste this resource, resulting
	in unnecessary dissipation.

	\section{Onward}
	
	The results of our analysis highlight the fundamental  thermodynamics limits of
	conventional computation---a limit we are steadily approaching but are still
	quite far from.  Constructively, our analysis also point to ways \emph{around}
	these limits for future hyper-efficient computers.  First, these hypothetical
	future computers could have implementations that adapt to the input biases and
	thus eliminate needless dissipation within each modular computation.  Second, we
	propose that future hyper-efficient computers can compute common composite
	routines in a single global transformation, to reduce modular dissipation.

\end{document}